\documentclass[11pt,dvips]{article}
\makeatletter
\renewcommand{\section}{\@startsection{section}{1}{0in}
{0.4\baselineskip}{0.1\baselineskip}{\Large\bf}}
\renewcommand{\subsection}{\@startsection{subsection}{2}{0in}
{0.25\baselineskip}{-\baselineskip}{\large\bf}}
\renewcommand{\subsubsection}{\@startsection{subsubsection}{3}{0in}
{0.1\baselineskip}{-\baselineskip}{\normalsize\bf}}
\makeatother

\begin{document}

%
%
%
{\it OG 3.3.25}

\begin{center}
%
%
{\Large {\bf Cosmic Ray Acceleration by Magnetic Traps}}
\footnote {Published in Proc. 26th ICRC 4, 439, Salt Lake City, USA, 1999}
\end{center}

%

\begin{center}
%
%
%
{\bf V.N.Zirakashvili$^{1}$}\\[0pt]\textit{$^{1}$Institute for Terrestrial
Magnetism, Ionosphere and Radiowave Propagation,Russian Academy of Sciences
(IZMIRAN), 142190, Troitsk, Moscow Region, Russia}
\end{center}

%

\begin{center}
{\large {\bf Abstract\\[0pt]}}
\end{center}

\vspace{-0.5ex}
%
%
%
Cosmic ray acceleration in turbulent interstellar medium is considered.
Turbulence is treated as ensemble of moving magnetic traps. We derive
equations for particle momentum distribution function that describes
acceleration of particles in this case. Rate of acceleration calculated is
estimated for our Galaxy and compared with ones given by other acceleration 
mechanisms.

%
\vspace{1ex}

%
%
%

\section{Introduction.}

Several types of distributed cosmic ray acceleration were investigated by many
authors: resonant stochastic acceleration (Tversko\u{i},1968), stochastic
acceleration by large scale turbulence (Bykov \& Toptygin, 1979), friction
acceleration (Berezhko, 1981, Earl, Jokipii,\& Morfill, 1988, Webb, 1989) and
in earlier papers of Fermi (Fermi, 1949) and Alfv\'en (Alfv\'en \&
F\"althammar, 1963). The last author introduced so called ''magnetic
pumping''. For such an acceleration mechanism particle changes its
perpendicular momentum in varying in time magnetic field, the gain of energy
being redistributed due to scattering. In this paper we shall introduce
further development of ``magnetic pumping'' and apply it for cosmic ray
acceleration by magnetohydrodynamic (MHD) turbulence.\newline In our Galaxy
interstellar MHD turbulence is mainly created by supernova (SN) explosions.
There exists turbulent transfer of kinetic and magnetic energies from the main
scale $L\sim100$ pc to smaller scales (Ruzmaikin, Sokolov,\& Shukurov, 1988).
In this paper we shall treat this turbulence as large amount of moving
magnetic traps. Energetic particles are trapped inside each trap and change
energy due to trap deformation. Weak scattering slowly changes pitch-angle of
the particle, determines escape time and hence rate of acceleration.

\section{Acceleration of particles in magnetic traps.}

Assuming that particles gyroradii are much smaller than magnetic trap size we
will use drift equation with scattering (Chandrasekhar, 1960)
$$
\frac{\partial f}{\partial t}+({\bf u}_{E}+v\mu{\bf b})\nabla
f+\frac{1-\mu^{2}}{2}\frac{\partial f}{\partial\mu}\left(  v\nabla
{\bf b}+\mu \nabla{\bf u}_{E}+3\mu{\bf u}_{E}({\bf b}%
\nabla){\bf b}\right)  -
$$
\begin{equation}
-p\frac{\partial f}{\partial p}\left(  \frac{1-\mu^{2}}{2}\nabla{\bf u}%
_{E}+\frac{1-3\mu^{2}}{2}{\bf u}_{E}({\bf b}\nabla){\bf b}\right)
=St_{u}f
\end{equation}

Here $f$ is cosmic ray momentum distribution function that is averaged on
gyroperiod, $p$ is particle momentum, ${\bf b}={\bf B}/B$ is the unit
vector along the magnetic field ${\bf B}$, $\mu={\bf pB}/pB$ is the
cosine of pitch angle of the particle, ${\bf u}_{E}=c[{\bf E\times
B}]/B^{2}$ is the electric drift velocity. For frozen magnetic field it is
simply perpendicular to ${\bf B}$ component of medium velocity ${\bf u}%
$. The right hand side of equation (1) describes scattering of particles by
small scale magnetic inhomogeneities moving with medium velocity ${\bf u}$.
Isotropic function $f_{o}$ that is independent on time and space coordinates
is exact solution of Eq. (1) for ${\bf u}=0$. Small deviation $\delta f$ of
cosmic ray momentum distribution function arises for nonzero ${\bf u}$.
Linearizing Eq.(1) we have
$$
\frac{\partial\delta f}{\partial t}+v\mu({\bf b}\nabla)\delta f
+\frac{1-\mu^{2}%
}{2}v\frac{\partial\delta f}{\partial\mu}\nabla{\bf b}-\frac{\partial
}{\partial\mu}\nu\frac{1-\mu^{2}}{2}\frac{\partial\delta f}{\partial\mu}=
$$
\begin{equation}
=p\frac{\partial f_{o}}{\partial p}\left(  \frac{1-\mu^{2}}{2}\nabla
{\bf u}_{E}+\frac{1-3\mu^{2}}{2}{\bf u}_{E}({\bf b}\nabla
){\bf b}\right)  \label{2}%
\end{equation}
Left hand side of Eq.(2) describes scattering and movement of particle, the
third term being correspond to magnetic mirroring. The terms proportional to
$\nu u$ are neglected since we are going to find solution of this equation in
the limit of small scattering frequency when the particle free path length
$\lambda=v/\nu$ is much greater than the trap size $l$. In this case it is
convenient to use new variable $q=(1-\mu^{2})/B$ instead of $\mu$. Let $\delta
f=\delta f^{+}$ for $\mu>0$ and $\delta f=\delta f^{-}$ for $\mu<0$, then
assuming quasistationarity of Eq.(2):
\begin{equation}
\pm{v}\frac{\partial\delta f^{\pm}}{\partial s}-2\frac{\partial}{\partial
q}\frac{\nu q}{B(s)}\sqrt{1-qB(s)}\frac{\partial\delta f^{\pm}}{\partial
q}=\frac{\frac{p}{2}\frac{\partial f_{o}}{\partial p}}{\sqrt{1-qB(s)}}\left[
qB(s)\nabla{\bf u}_{E}-(2-3qB(s)){\bf u}_{E}({\bf b}\nabla
){\bf b}\right]  \label{3}%
\end{equation}
Here $s$ is the coordinate along the magnetic field line. Our aim is to
perform averaging of Eq.(1) on ensemble of realizations of random magnetic
field and medium velocity. Since mirroring should be included the problem is
nonlinear. It will be simplified if one assumes that the medium velocity scale
is larger than the magnetic field scale and will treat them as statistically
independent quantities. We will also assume that medium is filled by many
identical magnetic traps, and magnetic field of each trap varies from
$B_{\min}$ up to $B_{\max}$. The same treatment was used for investigation of
influence of magnetic traps on the second harmonics of cosmic ray anisotropy
(Klepach \& Ptuskin, 1995).\newline For trapped particles $1/B_{\max
}<q<1/B_{\min}$. These particles perform oscillations inside a trap and slowly
diffuse on $q$. Hence function $\delta f^{\pm}$ for these particles is
approximately independent on $s$ and can be found from the equation:
\begin{equation}
-2\frac{\partial}{\partial q}\int\limits_{s_{1}(q)}^{s_{2}(q)}ds\frac{\nu
q}{B(s)}\sqrt{1-qB(s)}\frac{\partial\delta f^{\pm}}{\partial q}=\frac{p}%
{2}\frac{\partial f_{o}}{\partial p}\int\limits_{s_{1}(q)}^{s_{2}(q)}\frac
{ds}{\sqrt{1-qB(s)}}\left[  qB(s)\nabla{\bf u}_{E}-(2-3qB(s)){\bf u}%
_{E}({\bf b}\nabla){\bf b}\right]  \label{4}%
\end{equation}
Here $s_{1}(q)$ and $s_{2}(q)$ are roots of equation $B(s)q=1$. Performing one
integration on $q$ one obtains the following equation:
\begin{equation}
\int\limits_{s_{1}(q)}^{s_{2}(q)}ds\frac{\nu q}{B(s)}\sqrt{1-qB(s)}%
\frac{\partial\delta f^{\pm}}{\partial q}=\frac{p}{2}\frac{\partial f_{o}%
}{\partial p}\int\limits_{s_{1}(q)}^{s_{2}(q)}\frac{ds}{B(s)}\sqrt
{1-qB(s)}\left[  \frac{1}{3}\nabla{\bf u}_{E}(2+qB(s))+qB(s){\bf u}%
_{E}({\bf b}\nabla){\bf b}\right]
\end{equation}
Substituting the solution of this equation into Eq.(1), performing ensemble
averaging and integration on $\mu$ one can find momentum diffusion coefficient
(energy changes of trapped particles are taken into account only):
$$
D_{pp}=\frac{p^{2}w}{{4}}\left[  \int\limits_{0}^{l}\frac{ds}{B(s)}\right]
^{-1}\int\limits_{1/B_{\max}}^{1/B_{\min}}dq\left[  \int\limits_{s_{1}%
(q)}^{s_{2}(q)}ds^{\prime}\frac{\nu q}{B(s^{\prime})}\sqrt{1-qB(s^{\prime}%
)}\right]  ^{-1}\cdot
$$
\begin{equation}
\cdot\left\langle \left[  \int\limits_{s_{1}(q)}^{s_{2}(q)}\frac{ds}%
{B(s)}\sqrt{1-qB(s)}\left[  \frac{1}{3}\nabla{\bf u}_{E}%
(2+qB(s))+qB(s){\bf u}_{E}({\bf b}\nabla){\bf b}\right]  \right]
^{2}\right\rangle
\end{equation}
Here \ $w$ is the volume factor of traps and $\left\langle ...\right\rangle $
means ensemble averaging on ${\bf u}$. Expression (6) is rather cumbersome.
In order to simplify it we assume that the magnetic field disturbances are
small: $\delta B/B<<1$. In this case trapped particles have pitch angles close
to $\pi/2$, that is $\mu<<1$ and one can let $qB(s)=1$ in square brackets of
last integral. We will also neglect by second term in this integral in
comparison with first one. Assuming also that medium velocity scale is larger
than trap size $l$ and using $\mu$-independent scattering frequency one can
obtain
\begin{equation}
D_{pp}=\frac{p^{2}}{6}w\left\langle (\nabla{\bf u}_{\perp})^{2}%
\right\rangle \frac{1}{l\nu}\int\limits_{0}^{l}ds\sqrt{\left(  1-\frac
{B(s)}{B_{\max}}\right)  ^{3}}\label{7}%
\end{equation}
The deviation of momentum distribution function is given by the expression:
\begin{equation}
\delta f=p\frac{\partial f_{o}}{\partial p}\frac{\nabla{\bf u}_{\perp}%
}{2\nu}\left(  \frac{B_{\max}-B(s)}{B_{\max}}-\mu^{2}\right)  \theta\left(
\frac{B_{\max}-B(s)}{B_{\max}}-\mu^{2}\right)  \label{8}%
\end{equation}
where $\theta(x)$ is step function and ${\bf u}_{\perp}$ is perpendicular
to the mean field component of the medium velocity.\newline The expression (7)
can be readily obtained from the following estimation. For trapped particles
$\left|  \mu\right|  <\mu_{\ast}$. If trap is compressed or expands in
perpendicular to the mean field direction particle changes its momentum with
the rate $\frac{\Delta p}{\Delta t}\sim p\nabla{\bf u}_{\perp}$. The escape
time is $\Delta t\sim\mu_{\ast}^{2}/\nu$. Taking into account that trapped
particles are only $w\mu_{\ast}$ part of all particles one can find that
\begin{equation}
D_{pp}\sim\frac{(\Delta p)^{2}}{\Delta t}\sim p^{2}\frac{\left\langle
(\nabla{\bf u}_{\perp})^{2}\right\rangle }{\nu}w\mu_{\ast}^{3}\label{9}%
\end{equation}
If magnetic field disturbances have component parallel to the mean field
(magnetosonic wave magnetic disturbances) then $\mu_{\ast}\sim\sqrt{\delta
B/B}$ and we obtain expression (7).

\section{Discussion.}

It should be mentioned that expression (9) has the same form as the one for
the friction acceleration (Berezhko, 1981, Earl, Jokipii \& Morfill, 1988,
Webb, 1989). The friction acceleration was derived for the case when medium
velocity scale is larger than particle free path length. The acceleration
considered here works for opposite relation $l<<\lambda$ and thus can be
stronger for sufficiently small $l$. The same is concerned for comparison with
stochastic Fermi acceleration because it is of the same order as friction
acceleration for $l\sim\lambda$. The rate of resonance acceleration is roughly
(cf. Berezinsky et al.,1990)
\begin{equation}
D_{pp}^{res}\sim\Omega\frac{{v}_{a}^{2}}{{v}^{2}}\left(  \frac{\delta B}%
{B}\right)  ^{2}p^{2}\label{10}%
\end{equation}
where $\Omega$ is particle gyrofrequency, $v_{a}$ is Alfv\'{e}n velocity, and
$\delta B$ is the value of random magnetic field in the scale of the order of
particle gyroradius $r_{g}=v/\Omega$. In order to compare estimations (9) and
(10) one should take into account that $u\sim v_{a}\delta B/B$, $\mu_{\ast
}\sim\sqrt{\delta B/B}$, $w\sim1$ and take the smallest possible $l\sim r_{g}$
when the drift approximation is marginally valid. Taking for the scattering
frequency value that is determined by resonance scattering $\nu\sim
\Omega\left(  \delta B/B\right)  ^{2}$ (cf. Berezinsky et al., 1990) and
assuming that velocity scale is of the order of the trap size one can find
that
\begin{equation}
D_{pp}\sim\Omega\frac{{v}_{a}^{2}}{{v}^{2}}\left(  \frac{\delta B}{B}\right)
^{3/2}p^{2}\label{11}%
\end{equation}
Therefore resonant acceleration is a factor $\left(  \delta B/B\right)
^{1/2}$ less effective than considered here trap acceleration. If scattering
frequency drops at small $\mu$ the difference will be larger in favor of trap
acceleration. \newline It is well known now that resonant acceleration is
enough for weak reacceleration of galactic cosmic rays with energies about
1GeV/nucleon in order to explain mean thickness dependence at these energies
(Seo \& Ptuskin, 1994). Problems can arise for more powerful acceleration. It
seems that magnetosonic waves with lengths of the order of 1 GeV proton
gyroradius cannot exist in the interstellar medium because strong Landau
damping (acceleration rate estimation is obtained for such waves). \newline In
our Galaxy compressible turbulence exists in the main scale $L\sim100pc$ and
possibly at smaller scales due to nonlinear cascade transfer. The minimum
scale of compressible turbulence is determined by dissipative processes and is
estimated as $L_{\min}\sim0.01-1pc$ (Ruzmaikin, Sokolov,\& Shukurov, 1988). It
should note that this minimum scale can be also determined by dissipation
related with the acceleration mechanism considered.\newline The acceleration
rate (9) is determined by values of velocity and magnetic perturbations for
this minimum scale. In this case mechanism of acceleration considered has very
attractive feature: for ultrarelativistic particles the rate of acceleration
is proportional to space diffusion coefficient. Such a feature gives a
possibility of formation of power-low cosmic ray spectrum. This dependence is
valid up to the energies when particle gyroradius is approximately equal to
minimum scale of turbulence and is estimated as $Z\cdot10^{15}eV$ where $Z$ is
the particle charge, $L_{\min}\sim0.1pc$ and $B\sim10^{-5}Gs$ was assumed. The
ratio of diffusive exit time to acceleration time can be estimated as
\begin{equation}
\frac{\tau_{dif}}{\tau_{acc}}\sim\frac{H_{h}h_{acc}}{L_{\min}^{2}}\frac
{v_{a}^{2}}{v^{2}}\left(  \frac{\delta B_{\min}}{B}\right)  ^{7/2}\label{12}%
\end{equation}
Here $H_{h}$ is the galactic halo height, $h_{acc}$ is the height of
acceleration region, $\delta B_{\min}$ is the value of magnetic field
perturbation for minimum turbulence scale $L_{\min}$. This ratio is about 0.1
for $L_{\min}\sim0.1pc$, $H_{h}\sim10kpc$, $h_{acc}\sim3kpc$, $v_{a}%
\sim100km/s$, $\delta B_{\min}/B\sim1/10$. \newline For rough comparison with
acceleration on SN shocks one should note that both mechanisms use such a
source of energy as mechanical energy of SN explosion. Since only about 5
percent of initial mechanical energy of SN explosion goes into mechanical
energy of interstellar medium (Ruzmaikin, Sokolov \& Shukurov, 1988) one can
expect that if more than 5 percent of the initial mechanical energy of SNs
goes into accelerated at SN's shock particles, then acceleration mechanism
considered here is negligible in comparison with SN shock acceleration. In the
opposite case it can be important for our Galaxy.\newline
{\bf Acknowledgment.} This work was supported by RFBR (98-02-16347), INTAS
(95-16-711-23), ``Astronomy'' (1.3.8.1) grants and also by the Grant for young
scientists of Russian Academy of Sciences. V.N.Z. thanks the National Advisory
Committee for financial support for attending of 26th ICRC.

%
%
%
%
\vspace{1ex}

\begin{center}
{\Large {\bf References}}
\end{center}
Alfv\'{e}n, H. \& F\"{a}lthammar, C., 1963, Cosmical electrodynamics,
Oxford, Claredon press\\
Berezhko,E.G. 1981, Proc. 17th ICRC (Paris) 3,506.\\
Berezinsky, V.S., Bulanov, S.V., Dogiel, V.A., Ginzburg,
V.L.,\& Ptuskin, V.S., 1990,Astrophysics of Cosmic Rays, North-Holland Publ.
Comp.\\
Bykov, A.M. \& Toptygin, I.N. 1979, Proc. 16th ICRC (Kyoto)
2,66.\\
Chandrasekhar,S. 1960, Plasma Physics. The Univ. Chicago
Press.\\
Earl, J.A., Jokipii, J.R.,\& Morfill,G.E. 1988, ApJ
331,L91.\\
Fermi, E. 1949, Phys. Rev. 75, 1164.\\ 
Klepach, E.G. \&
Ptuskin, V.S. 1995, Bull. Russian Ac. of Sciences 59,672.\\
Tversko\u
{\i}, B.A. 1968, Sov. Phys.-JETP 26,821.\\
Ruzmaikin, A.A., Sokolov,
D.D., \& Shukurov, A.M. 1988, Magnetic Fields of Galaxies, Kluwer,
Dordrecht.\\
Seo, E.S. \& Ptuskin, V.S. 1994, ApJ 431,705.\\
Webb,
G.M. 1989, ApJ 340,1112.
\end{document}